# Mamba2 Meets Silence: Robust Vocal Source Separation for Sparse Regions

Euiyeon Kim and Yong-Hoon Choi
*Division of Robotics*
*Kwangwoon University*
Seoul, Korea
dmldussla93@gmail.com, yhchoi@kw.ac.kr

***Abstract*—We introduce a new music source separation model tailored for accurate vocal isolation. Unlike Transformer-based approaches, which often fail to capture intermittently occurring vocals, our model leverages Mamba2, a recent state space model, to better capture long-range temporal dependencies. To handle long input sequences efficiently, we combine a band-splitting strategy with dual-path architecture. Experiments show that our approach outperforms recent state-of-the-art models, achieving a cSDR of 11.03 dB—the best reported to date—and achieving substantial improvements in uSDR. Moreover, the model exhibits stable and consistent performance across varying input lengths and vocal occurrence patterns. These results demonstrate the effectiveness of Mamba-based models for high-resolution audio processing and open up new directions for broader applications in audio research. The code is available at https://github.com/EuiYeonKim/BSMamba2.**

## I. Introduction

Music source separation aims to isolate individual sources such as vocals, bass, drums, and guitar from a mixture, supporting applications in remixing, music information retrieval, and music education. Among audio separation tasks, it is particularly challenging due to the high sampling rate requirement (44.1 kHz), making advances in this area particularly valuable for other high-resolution audio domains. In this work, we focus on vocal separation, as vocals are the most critical element for conveying emotion and meaning in music, strongly influencing listeners' perception. High-quality vocal separation also enables the construction of datasets for singing voice synthesis, where isolated vocals from original tracks have been used to improve multilingual and multi-singer systems [1, 2], and provides a stronger basis for music transcription by enhancing melody extraction and score generation [3-5].

Recent years have seen rapid progress in music source separation, as reflected in the music demixing challenge (MDX) 2021 [6] and 2023 [7] challenges, which introduced state-of-the-art models such as HT Demucs [8] and BS-RoFormer [9]. Early approaches, including MMDenseNet [10], LaSAFT [11], ResUNetDecouple+ [12], and HT Demucs, were primarily U-Net based and operated on full spectrograms, but they struggled to capture fine-grained frequency details. Band-splitting methods such as BSRNN [13] later demonstrated effectiveness at high sampling rates by modeling inter-band dependencies, while BS-RoFormer further advanced performance by replacing recurrent neural networks (RNNs) with Transformers enhanced by rotary embeddings. However, BS-RoFormer struggles when vocals appear intermittently, since its global attention distributes focus uniformly across the sequence, failing to emphasize sparse but important vocal tokens. Prior work [14] suggests that selective attention to context-relevant tokens is more effective than uniform allocation in long sequences, motivating a reconsideration of model architectures for this task.

State space models (SSMs) have recently emerged as powerful alternatives for sequence modeling. Mamba [15] has shown strong results across domains, including speech enhancement and speaker separation, with variants such as SEMamba [16], SPMamba [17], and DPMamba [18] achieving state-of-the-art performance. Its successor, Mamba2 [19], introduces structural improvements that enhance both efficiency and long-sequence modeling. A notable feature of Mamba2 is selective state updates, which inject information strongly at important moments while suppressing irrelevant ones, making it particularly suitable for sparse-event sequences. Motivated by these insights, we propose *BSMamba2*, a model that combines band-splitting with the Mamba2 architecture to process high-resolution audio effectively. BSMamba2 addresses the limitations of Transformer-based approaches by capturing intermittent vocals more reliably and achieves robust, consistent separation performance across diverse conditions.

## II. Related Work

### *A. State Space Model*

State space models are dynamic systems that represent sequences in a latent state space. A standard SSM can be written as a linear dynamical system:

$$\mathbf{h}_t = \mathbf{A}\mathbf{h}_{t-1} + \mathbf{B}\mathbf{x}_t, \quad \mathbf{y}_t = \mathbf{C}\mathbf{h}_t, \tag{1}$$

where $\mathbf{h}_t$ denotes the hidden state at time, $\mathbf{x}_t$ is the input, and $\mathbf{y}_t$ is the output. The matrices $\mathbf{A}$, $\mathbf{B}$, $\mathbf{C}$ are learnable parameters that define the temporal dynamics of the system.

A representative approach is the hierarchical polynomial projection operators (HiPPO) framework [20], which projects input signals onto orthogonal polynomial function spaces (e.g., Laguerre or Legendre) and represents the sequence through the resulting polynomial coefficients. This formulation enables efficient modeling of long-range dependencies.

Building on SSMs, Mamba introduced the concept of selective SSMs, where system parameters are generated adaptively depending on the input sequence. Unlike conventional SSMs that share fixed $\mathbf{A}$, $\mathbf{B}$ across all timesteps, Mamba produces input-dependent parameters $\mathbf{A}_t$, $\mathbf{B}_t$, $\mathbf{C}_t$ allowing dynamic adaptation:

$$\mathbf{h}_t = \mathbf{A}_t\mathbf{h}_t + \mathbf{B}_t\mathbf{x}_t, \quad \mathbf{y}_t = \mathbf{C}^{\mathsf{T}}\mathbf{h}_t. \tag{2}$$

The matrices $\mathbf{A}_t$, $\mathbf{B}_t$ are discretized dynamically per timestep based on the input:

$$\mathbf{A}_t = e^{\Delta\mathbf{A}}, \quad \mathbf{B}_t = (\Delta\mathbf{A})^{-1}(e^{\Delta\mathbf{A}} - \mathbf{I})\Delta\mathbf{B}, \tag{3}$$

where $\Delta$ is an input-dependent parameter that governs selective state updates. This mechanism enables Mamba to propagate or suppress information adaptively, depending on the importance of the input signal at each timestep.

Mamba2 further improves efficiency and compatibility with Transformer-based architecture. Its main structural innovation simplifies the state transition matrix $\mathbf{A}$ from a diagonal form (used in S4D [21] and Mamba) to a scalar multiple of the identity:

$$\mathbf{A} = a\mathbf{I}, \tag{4}$$

where $a$ is a learnable scalar. This simplification reduces parameter count, improves computational efficiency, and—most importantly—makes Mamba2 structurally compatible with Transformer ecosystems. As a result, Mamba2 can leverage existing Transformer optimizations, hardware accelerations, and software infrastructure while achieving 2–8× faster runtime and competitive or superior performance across tasks.

### B. Application of SSM

Recently, Mamba-based architectures have demonstrated strong performance across a range of audio signal processing tasks. One notable example is SEMamba, which applies Mamba to speech enhancement. On the VoiceBank-DEMAND dataset, SEMamba achieved superior performance compared to Transformer- and Conformer-based models, reaching a perceptual evaluation of speech quality (PESQ) score of 3.69, the current state of the art. This improvement was further facilitated by integrating perceptual contrast stretching (PCS).

Another representative model is SPMamba, designed for speaker separation. It has achieved state-of-the-art results on benchmark datasets such as WSJ0-2Mix, WHAM!, Libri2Mix, and Echo2Mix. On Echo2Mix, SPMamba reported a signal-to-distortion ratio improvement (SDRi) of 16.1 dB and a scale-invariant signal-to-noise ratio improvement (SI-SNRi) of 15.3 dB, substantially outperforming prior convolutional neural network (CNN)-, RNN-, and transformer-based models. Moreover, compared to recent models like TF-GridNet, SPMamba reduces computational complexity and parameter count by nearly half, highlighting its efficiency.

Finally, DPMamba extends Mamba with a dual-path structure for speaker separation. By combining bidirectional Mamba within the dual-path framework, it effectively captures both local and global dependencies. On WSJ0-2Mix, the largest variant (DPMamba-L) surpassed the previous SOTA, Mossformer2, achieving a record SI-SNRi of 24.4 dB. Importantly, this performance was obtained with fewer parameters than competitive Transformer and CNN-based models, underscoring its computational efficiency.

These successful applications provide strong motivation for our approach, which replaces the RoFormer component in BS-RoFormer with Mamba2.

## III. The Proposed Method

We address the problem of vocal source separation, where the goal is to isolate the vocal signal from a given mixture of music. The input mixture is first transformed into a complex spectrogram $\mathbf{X} \in \mathbb{C}^{T\times F}$ using the short-time Fourier transform (STFT), where $T$ and $F$ denote the time and frequency axes, respectively. The model then estimates a vocal mask spectrogram $\mathbf{M} \in \mathbb{C}^{T\times F}$. Source separation is performed by applying the element-wise product of $\mathbf{M}$ and $\mathbf{X}$, followed by the inverse STFT to reconstruct the time-domain vocal signal.

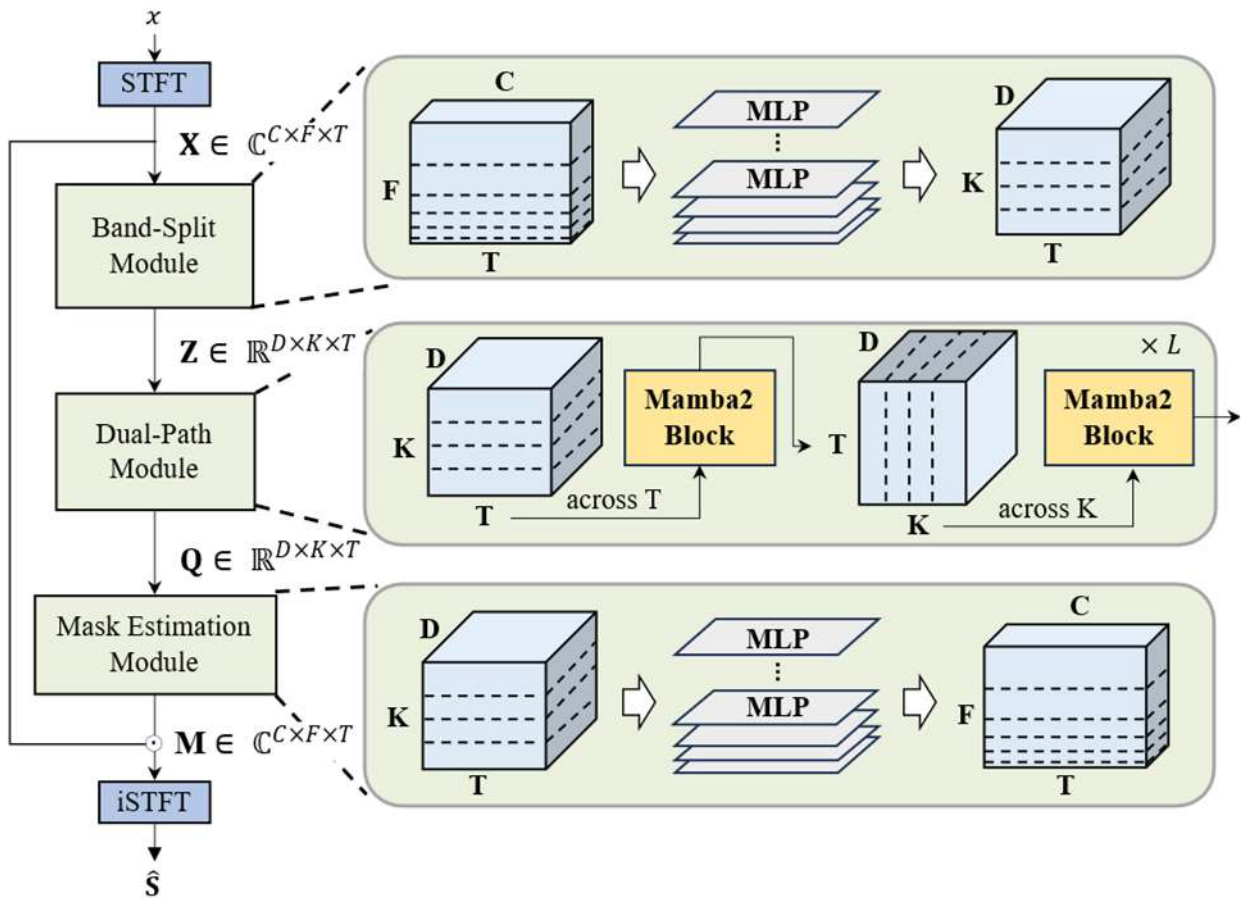

Fig. 1. BSMamba2 model architecture.

Our proposed BSMamba2 architecture follows the general design of BS-RoFormer and consists of three key components: (1) a band-splitting module, (2) a dual-path processing module, and (3) a mask estimation module. Figure 1 illustrates the overall structure of BSMamba2.

### A. Band-Split Module

The band-split module divides the input complex spectrogram $\mathbf{X} \in \mathbb{C}^{T\times F}$ along the frequency axis into $K$ sub-bands, and applies a separate feature extraction process to each sub-band. Each sub-band $\mathbf{X}_k$ corresponds to a unique frequency range $F_k$, with all sub-bands together covering the full frequency range, i.e., $\sum_{k=1}^{K} F_k = F$.

Each sub-band $\mathbf{X}_k \in \mathbb{R}^{T\times F_k}$ is mapped to a fixed-dimensional feature representation $\mathbf{Z}_k \in \mathbb{R}^{T\times D}$ through a multi-layer perceptron (MLP), where $D$ denotes the hidden dimension. The MLP consists of RMSNorm [22] followed by a linear layer and is applied independently to each sub-band. The resulting sub-band features are then stacked along the band axis to form the final feature tensor:

$$\mathbf{Z} = [\mathbf{Z}_1;\ \mathbf{Z}_2; \dots;\ \mathbf{Z}_K] \in \mathbb{R}^{T\times K\times D}. \tag{5}$$

### B. Dual-Path Module

The dual-path module is designed to effectively model sequential dependencies along both the time and sub-band axes. Given the input feature tensor $\mathbf{Z} \in \mathbb{R}^{T\times K\times D}$, it performs sequential processing in two directions. Bidirectional Mamba2 blocks are applied along each axis to enhance long-range dependency modeling.

Each bidirectional Mamba2 block consists of a forward and a backward Mamba2 block, with residual connections applied to their outputs, followed by a linear layer that merges the two directions and reduces dimensionality. Initially, $\mathbf{Z}$ is treated as independent sequences along the time axis $T$ for each sub-band and processed through the bidirectional Mamba2 block. This step captures temporal dependencies within each sub-band. Subsequently, the features are rearranged along the band axis $K$ and passed through another bidirectional Mamba2 block to model inter-band dependencies at each time frame. The dual-path processing is repeated $L$ allowing the model to progressively capture refined time–frequency interactions.

### C. Mask Estimation Module

The mask estimation module converts the sequential features $\mathbf{Q}$ obtained from the dual-path module into a time–frequency mask. The input features $\mathbf{Q} \in \mathbb{R}^{T\times K\times D}$ are first split along the sub-band axis $K$, and each sub-band $\mathbf{Q}_k \in \mathbb{R}^{T\times D}$ is processed independently.

Each sub-band feature $\mathbf{Q}_k$ is passed through an MLP consisting of RMSNorm for input normalization, two linear layers, an intermediate Tanh activation, and a final gated linear unit (GLU). The MLP outputs a mask $\mathbf{M}_k \in \mathbb{R}^{T\times F_k}$ corresponding to the frequency range $F_k$ of that sub-band. The sub-band masks are then concatenated along the frequency axis to form the final mask:

$$\mathbf{M} = [\mathbf{M}_1; \mathbf{M}_2; \dots; \mathbf{M}_K] \in \mathbb{R}^{T\times F}. \quad (6)$$

The resulting mask is applied element-wise to the input complex spectrogram $X$ to produce the estimated vocal spectrogram:

$$\hat{\mathbf{S}} = \mathbf{X} \odot \mathbf{M} \quad (7)$$

Finally, the inverse short-time Fourier transform (ISTFT) is applied to $\hat{\mathbf{S}}$ to reconstruct the time-domain vocal signal $\hat{\mathbf{s}}$.

### D. Loss Function

The model predicts the vocal source $\hat{\mathbf{s}}$, which is compared with the ground-truth vocal source $\mathbf{s}$ to compute the loss. We adopt the same loss formulation as BS-RoFormer, combining both time-domain and frequency-domain information. The total loss is defined as:

$$\mathcal{L} = \lambda_{\text{time}} \cdot \|\hat{\mathbf{s}} - \mathbf{s}\|_1 + \sum_{f\in\mathcal{F}} \left\| STFT_f(\hat{\mathbf{s}}) - STFT_f(\mathbf{s}) \right\|_1 \quad (8)$$

The first term is the L1 loss in the time domain, while the second is a multi-resolution STFT loss that aggregates the L1 distances between complex spectrograms at multiple resolutions. For the multi-resolution STFT loss, we use hop size 147 and window sizes of [4096, 2048, 1024, 512, 256]. The weighting factor for the time-domain loss is set as $\lambda_{\text{time}} = 10$.

## IV. Experiments

### A. Dataset

We conduct our experiments on the MUSDB18HQ dataset [23], a publicly available high-quality music source separation dataset sampled at 44.1 kHz in stereo. Each track contains isolated stems for vocals, drums, bass, and other instruments. The dataset consists of 150 tracks in total. Following the commonly used split, we use 86 tracks for training, 14 tracks for validation, and 50 tracks for testing.

### B. Metrics

We evaluate model performance using the signal-to-distortion ratio (SDR), a widely adopted metric in music source separation. SDR measures the degree of distortion between the predicted and ground-truth sources, with higher values indicating better separation quality. In this study, we report two types of SDR: chunk-level SDR (cSDR) and utterance-level SDR (uSDR). For cSDR, each test track is divided into 1-second audio chunks, and SDR is computed for each chunk. The median SDR is taken for each track, and the final score is obtained by computing the median across all tracks. For uSDR, SDR is calculated over the entire track, and the mean across all tracks is reported as the overall model performance.

TABLE I. Hyperparameter settings for BS-RoFormer and BSMamba2.

| Hyperparameters | BS-RoFormer | BSMamba2 |
|---|---|---|
| Hidden dimension $D$ | 384 | 256 |
| Dual-path modules $L$ | 6 | 6 |
| Number of sub-bands | 62 | 62 |
| Learning rate | $5 \times 10^{-4}$ | $5 \times 10^{-4}$ |
| Precision | float16 | bfloat16 |
| Total parameters | 72.2M | 48.1M |
| Number of GPUs | 4 | 2 |
| Batch size per GPU | 4 | 5 |
| Gradient accum. steps | 4 | 6 |
| Effective batch size | 64 | 60 |

TABLE II. Performance comparison of vocal source separation. The best performance is **bold**.

| Model | uSDR | cSDR |
|---|---|---|
| ResUNetDecouple | – | 8.98 |
| Hybrid Demucs | – | 8.13 |
| BSRNN | 9.73 | 10.01 |
| SIMO stereo BSRNN [24] | 10.33 | 9.79 |
| BS-RoFormer(L=6, TC) | – | 10.68 |
| SCNet-large [25] | – | 10.86 |
| BS-RoFormer (unofficial) | 10.29 | 10.47 |
| BSMamba2 (ours) | **10.70** | **11.03** |

TABLE III. Comparison of uSDR performance across different input lengths.

| Model | 1–2 s | 2–4 s | 4–8 s |
|---|---|---|---|
| BS-RoFormer (unofficial) | 8.47 | 10.29 | 12.14 |
| BSMamba2 (ours) | **9.62** | **11.28** | **12.39** |

### C. Implementation Details

For a fair comparison, we re-implemented both the proposed BSMamba2 model and the baseline BS-RoFormer under identical training environments and dataset settings, keeping the architecture and hyperparameters of BS-RoFormer the same as in the original work. All input audio was segmented into 8-second clips with a sampling rate of 44.1 kHz, and complex spectrograms were computed using STFT with a window size of 2048 and hop size of 441. Mixtures were created by randomly combining the four sources (vocals, drums, bass, and other), and various data augmentation techniques were applied to enhance diversity.

Table I summarizes the hyperparameters used for both BS-RoFormer and BSMamba2. Training was conducted on NVIDIA A100 80GB GPUs. BS-RoFormer used four GPUs with a per-GPU batch size of 4 and gradient accumulation of 4, yielding an effective batch size of 64, closely matching the original setup. BSMamba2 used two GPUs with a per-GPU batch size of 5 and gradient accumulation of 6, resulting in an effective batch size of 60. During evaluation, model outputs were generated by sequentially concatenating non-overlapping 8-second segments to reconstruct the full audio.

TABLE IV. Vocal separation performance of BS-RoFormer and BSMamba2 across input lengths (1–16 Seconds). The best performance is **bold**.

| Model | 1 s | | 2 s | | 4 s | | 8 s | | 12 s | | 16 s | |
|---|---|---|---|---|---|---|---|---|---|---|---|---|
| | uSDR | cSDR | uSDR | cSDR | uSDR | cSDR | uSDR | cSDR | uSDR | cSDR | uSDR | cSDR |
| BS-RoFormer (unofficial) | 8.09 | 8.72 | 9.37 | 9.85 | 10.00 | 10.25 | 10.29 | 10.47 | 9.47 | 9.70 | 8.97 | 9.26 |
| BSMamba2 (ours) | **9.27** | **10.12** | **10.00** | **10.65** | **10.50** | **10.89** | **10.70** | **11.03** | **10.78** | **11.07** | **10.73** | **11.04** |

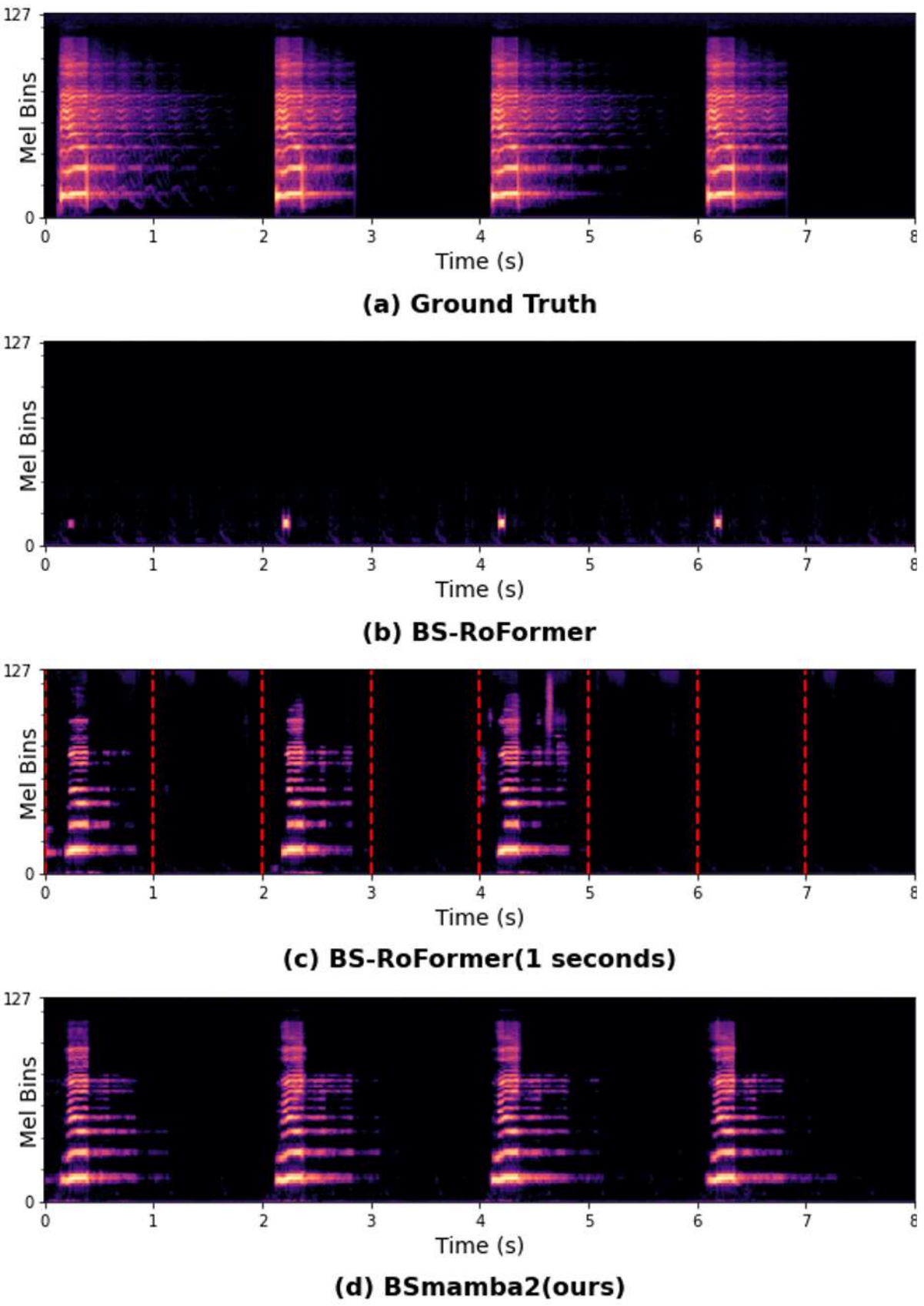

Fig. 2. Comparison of mel spectrograms generated by different models.

## V. Results

### A. Performance Comparison with Baseline Models

Table II presents the vocal separation performance of the proposed BSMamba2 compared with various recent source separation models. We report both uSDR and cSDR, with BSMamba2 achieving the best performance across both metrics.

Notably, BSMamba2 attains the highest performance among existing band-split models, surpassing the previous official state-of-the-art cSDR achieved by SCNet-large. Compared to the BS-RoFormer (unofficial) trained under the same conditions, BSMamba2 improves uSDR by +0.41 dB and cSDR by +0.56 dB. This demonstrates that simply replacing the RoFormer blocks with Mamba2 in the BS-RoFormer architecture leads to a clear enhancement in separation quality.

Furthermore, while other band-split models generally benefit from increased network depth, BSMamba2 improves performance without increasing the depth beyond that of BS-RoFormer. These results indicate that BSMamba2 offers a superior alternative in terms of both performance and scalability, even with a relatively simple structural modification.

### B. Separation Performance Analysis by Vocal Onset Time

Figure 2 shows the mel-spectrogram results for a segment of *PR-Happy Daze* from the MUSDB18 test set, comparing the ground truth, BSMamba2, BS-RoFormer, and 1-second chunk inference of BS-RoFormer in scenarios where vocals occur intermittently. Both BSMamba2 and BS-RoFormer used 8-second input chunks for inference, while the 1-second inference was obtained by sequentially concatenating eight output chunks. The 8-second inference of BS-RoFormer failed to clearly separate the vocal regions, whereas the 1-second inference demonstrated relatively improved separation. This suggests that in longer chunks, sparse vocal occurrences cause attention-based modules to insufficiently focus on vocal information, while shorter chunks provide more continuous vocal presence, mitigating this issue. In contrast, BSMamba2 produced clearer separation even under the same conditions, indicating that Mamba2 blocks capture intermittent vocals more effectively than attention-based modules.

Table III reports the separation uSDR performance of BS-RoFormer and BSMamba2 with respect to vocal onset duration. The performance gap was largest (1.15 dB) for short vocal segments of 1–2 seconds, while the difference decreased to 0.25 dB for longer segments of 4–8 seconds. This indicates that BS-RoFormer's performance deteriorates substantially when vocals appear intermittently, whereas BSMamba2 maintains more consistent separation quality.

### C. Performance Comparison Across Input Durations

Table IV presents the vocal separation performance of BS-RoFormer and BSMamba2 when varying the input length from 1 to 16 seconds. Both models were trained with 8-second input chunks, but the experiments evaluated not only shorter but also longer input durations. Around the 8-second training length, the performance gap between the two models was relatively small. However, as the input became shorter or longer, the difference widened. For longer inputs, BS-RoFormer's performance dropped sharply, whereas BSMamba2 effectively handled extended temporal contexts, even exhibiting a slight performance advantage.

For shorter inputs, the performance gap arises because BSMamba2 can accurately capture vocal characteristics with relatively limited information. This suggests that training on longer sequences could further enhance performance. In contrast, BS-RoFormer may appear effective for short segments due to fewer intermittent vocals, but the lack of sufficient contextual information can hinder understanding of the overall musical structure. Therefore, caution is required when interpreting results for short input lengths, as they can inadvertently reduce overall performance.

## VI. CONCLUSION

In this paper, we proposed BSMamba2, a novel model for accurately extracting vocal sources in music source separation. The model replaces the Transformer-based RoFormer modules in the existing BS-RoFormer architecture with the recently proposed Mamba2 modules, effectively addressing the limitations of previous models in separating intermittent vocal sources. As a result, BSMamba2 achieves stable and superior separation performance even when vocals occur sparsely. Experimental results demonstrate that BSMamba2 outperforms various recent state-of-the-art models, achieving a cSDR of 11.03 dB, surpassing the previous best performance, and showing substantial improvements in uSDR as well. Further analysis empirically confirms that BSMamba2 successfully mitigates the performance degradation of BS-RoFormer in sparse vocal scenarios.

## ACKNOWLEDGMENT

This work was supported by the Korea Agency for Infrastructure Technology Advancement under Grant RS-2025-02532980 funded by the Ministry of Land, Infrastructure and Transport under the Smart Building R&D Program; and by the National Research Foundation of Korea (NRF) grant funded by the Korea government (MSIT) (RS-2025-16069933).